*Reference: van Haren, H., H.A. Dijkstra, 2021. Convection under internal waves in an alpine lake. Env. Fluid Mech., 21, 305-316.*# Convection under internal waves in an alpine lake

by Hans van Haren[*1], Henk A. Dijkstra[2]

[1]Royal Netherlands Institute for Sea Research (NIOZ), P.O. Box 59, 1790 AB Den Burg, the Netherlands.
*e-mail: hans.van.haren@nioz.nl

[2]Institute for Marine and Atmospheric research Utrecht, Department of Physics, Utrecht University, Utrecht, the Netherlands.

**Abstract**

Turbulent mixing processes in deep alpine Lake Garda (I) have not extensively been observed. Knowledge about drivers of turbulent fluxes are important for insights in the transport of matter, nutrients and pollutants, in the lake and in natural water bodies in general. In this paper, the occurrence of internal-wave-induced turbulent convection, termed 'internally forced convection', is addressed as opposed to the more common shear-induced turbulence in a density stratified environment. Observations are analyzed from a dedicated yearlong mooring holding 100 high-resolution temperature sensors at 1.5 m intervals under a single current meter in the deeper half of the 344 m deep lake-center. Episodically, the weakly density stratified waters in the lower 50 m above the lake-floor show spectral slope and coherence evidence of short-term (15 to 30 minutes) convective motions under internal waves that are supported by the stronger stratified waters above. The near-homogeneous conditions are not attributable to frictional Ekman dynamics, but to large-scale internal wave crests.

# 1 Introduction

'Natural (free)' turbulent convection is generally considered to occur during a period of gravitational instability when denser (relatively cool) fluid is over less dense (warmer) fluid [1-3]. It is an efficient mixing process [4] and therefore important for fluxes of nutrients and suspended particles in geophysical environments like the atmosphere and ocean. Natural convection commonly occurs in the atmosphere due to solar radiation reflecting from the Earth surface below. In a bounded fluid, convection is classically studied in a laboratory oil-filled pan over a heat source as Rayleigh-Bénard convection [e.g., 5]. This convection starts at sufficiently high Rayleigh number $Ra = g\alpha\Delta T h^3/(\nu\kappa) > Ra_c > 1000$, where the critical $Ra_c$ depends on boundary conditions, g denotes the acceleration of gravity, $\alpha$ the thermal expansion coefficient, $\Delta T$ the vertical temperature difference driving the positive buoyancy, h the layer depth, and $\nu$ and $\kappa$ the kinematic viscosity and thermal diffusivity, respectively. It is characterized by up- and down-ward cells of motion that have an aspect ratio (of vertical over horizontal dimensions) of approximately unity [6]. In natural waters like oceans and lakes, $Ra = O(10^8)$ and the convective, full turbulence has more than one (overturning-)cell-size, in fact many from the large (>100 m) energy containing scales to the 0.001 m dissipation scales. Multiple scales of irregular structures are seen in the development of initial value Rayleigh-Taylor instabilities in a fluid interior between cold water overlying warm water [e.g., 7]. Upon rotation of the frame of reference, the spatial pattern of such convection consists of up- and down-ward motions in plumes of less dense and denser fluids at the largest scale that are irregular along the edges due to secondary and smaller-scale motions.

Natural convection drives the atmospheric boundary layer turbulence, mainly during daytime, and it is well-established in driving the nighttime (cooling) near-surface ocean and lake turbulence [e.g., 7, 9-11]. Near-surface convective mixing can also include that driven by nonlinear internal waves, especially in their core where the particle velocity exceeds the phase speed [12-15]. However, convection has not often been observed in the deep natural water bodies. This is because of the dominant vertical density stratification in the deep ocean due to the large daytime solar heating from above, in spite of geothermal heating from below [e.g., 16]. Apparently for geothermal heating (less than 1% of solar heating [17]) to be observed directly, one requires homogeneous waters near the ocean-floor and weak horizontal density gradients and horizontal currents. Near-homogeneity is a rare phenomenon in the ocean, and which requires high-resolution observations to be able to record, e.g., temperature differences smaller than 0.0001°C. In near-homogeneous waters under 'internal waves' that are supported by the stable density stratification above, occasional up- and downward turbulent motions have been observed that are reminiscent of convective turbulence [18]. As these motions were observed using temperature sensors moored above sloping topography in the ocean, a possible contribution of salinity besides temperature to density variations could not be established. As a result, convective turbulence could only be hypothesized.

Laboratory and numerical experiments have shown that natural convection only occur when gravity is overcome, either by a continuous unstable set-up at the boundary as in Rayleigh-Bénard convection, or by applying bulk forcing acceleration exceeding 1g of an entire container with a stably stratified fluid as in Rayleigh-Taylor instability [e.g., 19]. Internal waves usually do not accelerate over 1g, but they may overcome 1g', for reduced gravity $g' = g\Delta\rho/\rho$. If internal waves set-up convection in the near-homogeneous layer adjacent to their stratified layer, the rationale would be an internal forcing generation, not a bulk forcing, of natural convection in a stably stratified fluid [18]. We term this 'internally forced convection'.

In this paper, the phenomenon of internally forced convection is evidenced in substantial observational detail from high-resolution temperature sensors moored in Lake Garda, Italy. In this freshwater mountain lake salinity does not contribute significantly to density variations. The observations during a period dominated by turbulent convection are contrasted with those from a period with shear-dominated turbulence about one day later in the records. Shear and convection are evaluated in the time domain by comparison with previous near-surface convection observations [e.g., 12, 15] and direct numerical simulations of Rayleigh-Taylor instabilities [7], and in frequency (spectral) domains with specific distinctions in slope-scaling between shear and convection [20, 21].

The paper is organized as follows. In Section 2 the observational site, data and analysis methods are described. In Section 3 the observations are presented. In Section 4 suggestions for the internal forced convection are discussed.

**2 Site, data and methods**

Lake Garda is an approximately 40 km long and 5 km wide lake roughly directed north-south, with depths up to 346 m. For one year, a taut-wire sub-surface mooring was installed near its deepest point (Fig. 1), at latitude 45° 42.947´N (at which the local inertial frequency $f = 1.04 \times 10^{-4}$ s$^{-1}$, providing inertial period $T_f = 16.7$ h), longitude 10° 44.567´E, between 10 UTC on 24/05/17 (yearday 144.42) and 09 UTC on 31/05/18 (yearday 516.37; days in 2018 are +365). The local water depth is $H = 344$ m with a relatively weak bottom slope (100 m scale) of about $\beta = 3°$. At a distance of 300 m to the East and 1.3 km to the West, the lake-floor shallows with slopes varying between 10° and 20°.

The mooring consisted of floatation near its top providing 1100 N net buoyancy (Fig. 1). Below the floatation, a single point Nortek AquaDopp current meter was at $z = -187$ m, sampling current components [u v w] and storing ensemble data every 300 s. The current meters' tilt and pressure sensors demonstrated that the mooring-top did not move more than 0.05 m vertically and not more than 5 m horizontally, under maximum current speeds of 0.17 m s$^{-1}$. Typical current speeds vary between 0.01 and 0.05 m s$^{-1}$.

In the vertical range $-338 < z < -189.5$ m, 100 'NIOZ4' self-contained high-resolution temperature T-sensors were located at 1.5 m vertical intervals, sampling at a rate of once per 2 s. NIOZ4-sensors have a precision of better than 0.0005°C after drift correction, a noise level of less than 0.0001°C and a resolution of less than 0.00001°C [22]. The T-sensors are synchronized via induction every 4 hours and a vertical profile of 148.5 m is measured in less than 0.02 s.

Drift of nominally 0.001°C/mo for aged sensors is corrected by fitting typically four day mean temperature profiles to a smooth statically stable multiple order polynomial profile. The mean profile is tuned with local shipborne Conductivity, Temperature Depth (CTD-)profile data obtained roughly once every 6 weeks by the Environmental Protection Agency of the Veneto Region within 1 km from the mooring site. Up to day 350, 7 sensors showed various electronic (noise, calibration) problems and are not further considered. The measured temperatures are transferred to potential temperature θ to correct for pressure-compressibility [23]. As the observations were made in a freshwater lake with minimum temperatures of about 8 °C (>4 °C of maximum density), the temperature-density relationship is always tight. Salinity contributes at best a factor of 7 less than temperature to density variations, as has been verified with the CTD-data. Further details on the site and mooring details, together with an overview of the entire observational period can be found in [24].

The high-resolution temperature data will be presented as maps in depth-time. These maps are compared with previous results of Rayleigh-Taylor instability studies in laboratory and numerical simulations and convection in near-surface internal waves [12-15]. For example, direct numerical simulations [7] demonstrated more quasi-regular and fingerlike, slower Rayleigh-Taylor instability development when the entire fluid-system was rotating, compared with a non-rotating case.

The temperature data are also analysed in the frequency ($\sigma$) domain. It has been demonstrated [20] from analyses of higher order moments and structure functions that convection can be distinguished from shear-induced turbulence. The Gaussian value of fourth moment flatness is clearly exceeded only during convective turbulence of an active scalar, when the second moment spectral slope significantly deviates from the inertial subrange slope-scaling of $\sigma^{-5/3}$ representing shear turbulence of a passive scalar [25, 26]. While the spectral slope of convection has not been found previously in ocean observations [20], laboratory experiments [21] demonstrated that at least in the lower frequency part of inertial subrange a close match is found with $\sigma^{-7/5}$ or the Bolgiano-Obukhov scaling for an active scalar convective turbulence. Originally, this scaling was proposed for the buoyancy range of stratified turbulence, between the internal wave and inertial subranges [27]. We scale our spectra with the inertial subrange slope-scaling and compare with various deviating slopes, including +4/15, the difference between 5/3 and 7/5, +2/3 scaling of open-ocean internal waves [28], and -1/3 for fine-structure contamination [29].

## 3 Observations

A five-day overview of fixed point observations demonstrates a typical response in Lake Garda to a wind event blowing from the north (Fig. 2). Such events occur roughly every 10 days [14]. The response to the sudden wind increase on day 316 (Fig. 2a) is an almost instantaneous increase in mid-depth flow to the north, followed by a flow to the south commencing one inertial period after the peak to the north (Fig. 2b). With the flow to the south, the temperature stratification is depressed by several tens of meters (Fig. 2c). Upon the relaxation of the depression, a train of high-frequency internal waves occurs over a weakly stratified layer lasting approximately two inertial (16.7 h) periods, each inertial period demonstrating distinctly different turbulence in the lower layer above the lake-floor.

Rather unusual up- and downward motions underneath high-frequency internal waves are seen in magnifications of the first inertial period (Fig. 3). With the aid of transferring the horizontal time-axis into a spatial axis using the mean mid-depth current speed, we computed that the approximately 15 minute scale motions are roughly 20 m large. The portions of upward moving cooler water and downward moving warmer water thus have aspect ratio of about 1. They are irregular along the edges. With respect to images on (laboratory, modelling) observations of natural convection [e.g., 7], the images in Fig. 3 are upside-down, as the instabilities are slightly larger at the warmer, stable side. All characteristics of irregular, skewed up- and down-going motions point at convective, not shear-induced, turbulent overturning, likely under influence of rotation given the columnar structure and demonstrating more fine-scales than in simulations [7]. However, the observed convection occurs in a larger-scale stable stratification under internal wave action. We recall that the present freshwater Lake Garda temperature data in Fig. 3 are undisputed tracers for density variations, unlike observations from the ocean where salinity also may substantially contribute to density variations. Below, energy spectra and coherence are used to qualitatively support time series analyses on the two contrasting turbulence processes of more common shear-induced (Fig. 4) and relatively rare convective (Fig. 5) overturning under internal waves.

**3.1 Correspondence between the two inertial periods**

Both inertial periods have in common that the larger stratification (Fig. 4a, 5a) in the upper 45 m layer of the sampled range is reflected in larger temperature variance compared to that of the lower near-bottom layer (Fig. 4d, 5d). In the second half of each image (Fig. 4a, 5a), both inertial periods demonstrate a weakly (minimum) stratified layer with mean $N = 2f$ that lasts about half an inertial period and which extends about 40 to 50 m above the lake-floor, while varying over shorter timescales. The average Ozmidov scale of largest overturn in a stratified environment is smaller for the upper layer than for the lower layer, being smaller than and larger than the 1.5 m separation distance between the T-sensors, respectively. Both inertial periods show a gradual decrease in coherence for frequencies $\sigma > N$ at all depths (Fig. 4e, 5e). Both inertial periods also show less coherence in the well-stratified upper layer compared to the weakly stratified lower layer, not only at the smallest 1.5 m scale displayed in Fig. 4e, 5e, but also at larger scales, for all $\sigma > N$. Details however, vary between the two inertial periods that are attributable to different predominant turbulence processes in the lower layer.

**3.2 Differences between the two inertial periods**

The inertial period in Fig. 4 shows less hair-like variations in temperature (stratification) in the lower layer up to 50 m above the lake-floor (Fig. 4a,b), up to one order of magnitude less variance (Fig. 4d), and less coherence (Fig. 4e), compared with the inertial period of Fig. 5. In the upper layer in more stratified waters of Fig. 4, the temperature variance shows a broad peak around the first inertial higher harmonic 2f. This peak is absent in Fig. 5d. If associated with near-inertial waves, as a first harmonic, they may be associated with near-inertial shear. Such a shear is the dominant shear in stratified seas because of the short vertical length scales of near-inertial waves [e.g. 30]. This potential of near-inertial shear may explain the near-lake-floor spectrum being close to the $\sigma^{-5/3}$-scaling of inertial subrange turbulence of a passive scalar in Fig. 4d [25, 26], throughout the range $\sigma > N_{max}$, the local maximum small-(1.5 m) scale buoyancy frequency.

The inertial period in Fig. 5, with overall mean turbulence level of about twice that of Fig. 4 [24], is characterized by a growing of the weakly turbulent layer near the lake-floor with apparent convection type up- and downward turbulence motions of which the scales have an aspect ratio of about 1 (cf., Fig. 3 which is a magnification of Fig. 5c). The 20-m, 100 cycles per day (cpd) scale motions are indeed found significantly coherent at the 1.5 m T-sensor separation (Fig. 3e). However, much higher frequency motions are still found coherent (at the 1.5 m scale), up to about 2000 cpd.

The corresponding temperature variance spectrum, which is more energetic over a broad range $\sigma > 2f$ than in Fig. 4, does not follow the $\sigma^{-5/3}$ scaling law of shear-induced turbulence (Fig. 5d). For $\sigma < 1000$ cpd the spectral slope (on the log-log plot) has a positive slope with respect to the $\sigma^{-5/3}$ scaling law, while for $\sigma > 1000$ cpd it has a negative slope with respect to the $\sigma^{-5/3}$ scaling law. The different spectral slopes in Fig. 4d and Fig. 5d distinguish convection from shear dominance in turbulent overturning [20, 21]. The significant deviations from the inertial subrange $\sigma^{-5/3}$ scaling point at convective turbulent overturning of an active scalar. They are seen to scale as $\sigma^{-7/5}$ over a two orders of magnitude frequency range including the Ozmidov-scale for the lower layer in Fig. 5. This is a larger portion of the inertial subrange than found in laboratory experiments [21]. The portion $\sigma > N$ of the upper layer (high variance, stronger stratified environment) spectrum is closer to the $\sigma^{-5/3}$ scaling of a passive scalar, as for both layers in Fig. 4d.

**4 Summary and discussion**

It remains to be discussed why internally forced convection is mainly observed in the lower 50 m above the lake-floor when high-frequency internal wave activity is rather intense in Figs 3,5. The weakly stratified layer near the lake-floor cannot be generated by <0.1 m s$^{-1}$ frictional flows, which would yield a near-homogeneous layer of only a few meters thick [31]. Several possibilities exist for internal waves to generate the weakly stratified conditions and convective turbulence over vertical ranges of several tens of meters above the lake-floor, as sketched in Fig. 6.

One suggestion is that propagating high-frequency internal waves strain the stratification, which results in alternate phases with larger than average stratified waters in a thin layer close to the lake-floor following downward motions with less than average stratification following upward motions (Fig. 6a). This is reminiscent of convective processes in the core of near-surface nonlinear solitary internal waves where the particle velocity exceeds the phase velocity [13], notably after interaction with high-frequency internal waves [32, 33]. Near the lake-floor, the upward motion may serve as a preconditioning of sufficiently weakly stratified waters in which convection can penetrate during the downward phase. The solid boundary of the lake-floor acts as a one-sided restricting limit for the straining of isopycnals.

An alternative suggestion is inspired by Thorpe [34, 35] who proposed internal wave breaking patches with convective overturning in obliquely propagating internal waves in the ocean interior (Fig. 6b). In a similar fashion, nearby sloping topography as in the lake's sidewalls may favour strongly non-linear internal wave propagation and associated patches of, possibly intensified, convective turbulence.

Such convective instabilities under obliquely propagating (non)linear internal waves may interact with rotational physics processes. For example, McEwan [36] suggested 'inertial' gyroscopic wave breaking providing columnar vortices and, possibly, yielding slanted convection [37, 38]. The convection will be, for pure homogeneous water, in the direction of the Earth rotational vector. At mid-latitudes, this will appear as weakly stratified water ($N = f$, 2f, 4f, depending on the prevailing stability mechanism [39]). In the present observations, $N = 2f$ dominates in the weakly stratified layer near the lake-floor, which suggests rotational effects are important and explains why time-depth images like Fig. 3 correspond best to Rayleigh-Taylor instability simulations that include rotation [7].

The observations on convective mixing under internal waves in the deeper weakly stratified part of Lake Garda demonstrate the richness of turbulence generation phenomena in geophysical environments. It is noted that in these observations temperature is a unique tracer for density variations.

It seems that the accelerations of the observed convective turbulence, in combination with the internal wave vertical motion variations, can overcome reduced gravity to sufficiently penetrate the stable weakly stratified layer below. Reduced gravity,

$$g' = g\Delta\rho/\rho \approx 2gA,$$

is the natural driving force for internal waves in stable-stratified fluids. The Atwood number,

$$A = \Delta\rho/\Sigma\rho,$$

denotes the ratio of the density ($\rho$) difference across an interface and the sum of the different densities in both layers above and below the interface [3]. In Lake Garda, $A \approx -10^{-7}$ which is roughly calculated following the observations in Figs 3, 5 of $\Delta T \approx 0.003$°C, with an associated $\Delta\rho$ of 0.0003 kg m$^{-3}$ [24], and it is proposed that g' should replace g in considering accelerations to overcome stable stratification [18]. As the present wave-induced penetration has vertical motion variations of $\Delta w \approx 0.01$ m s$^{-1}$, roughly observed by the current meter near mid-depth assuming little vertical 'phase' variation, the condition for the magnitude of its accelerations is,

$$|\Delta w/\Delta t| = a_w > g',$$

which is satisfied when $\Delta t < 3000$ s. This time-variation limit of up- and down-going motions is close to the observed minimum buoyancy period in Figs 3, 5, or the largest turbulence timescale of the entrainment overturns observed. The correspondence with the minimum buoyancy period is similar to ocean observations [18], in which values were an order of magnitude different from the ones presented here.

The overcoming of reduced gravity in internally forced convection by internal waves as opposed to gravity in bulk forcing of an entire water basin is rationalized in analogy of parcel displacement leading to the definition of buoyancy frequency [40]. After all, in a stably stratified environment the vertical displacement of a parcel of density $\rho$ yields a restoring force g' that sets high-frequency internal waves into motion at frequency N.

The internally forced convection of a stratified fluid has a peculiar characteristic in that when it is forced at $a_w$ > g', it will reinforce itself by its secondary shear-induced mixing that entrains heavier fluid in a downward going column that is less dense than the environment. This reduces the density difference between column and environment, until the difference disappears at the deepest point of the column. This process is at odds with natural convection, which requires continuous (bulk) forcing by gravity as a downward going denser column will entrain less dense fluid from its statically unstable environment, thereby counteracting the convective motion. Semantically, internally forced convection is more natural.

Further laboratory and numerical modelling is required to test the above suggestions and which may best explain the presented observations of internally forced convection in weakly stratified waters under internal waves of different frequency. One could create two different excitations of a stratified rotating fluid bounded by weaker stratification below, one exciting near-buoyancy frequency waves, the other near-inertial waves. Or, a large depression is excited and the development is studied in a stratified rotating fluid, in which then also the removal of high-frequency waves may be studied at the expense of near-inertial waves.

**Acknowledgements** We highly appreciated the assistance of the Nautical Rescue Team 'Vigili del Fuoco-Trento', S. Piccolroaz and M. van Haren around the deployment and recovery of the mooring. We thank M. Laan for all his temperature sensor efforts. S. Piccolroaz supplied the map of Lake Garda and mooring scheme. This project was funded by the Faculty of Science of Utrecht University through a grant to HD and was supported by the National Marine Facilities (NIOZ).

**Fig. 1**. (a) Central part of Lake Garda in Northern Italy with bathymetry, coordinate axes and mooring location (red star) near the lake's deepest point. (b) The instrumented mooring line of about 170 m between top-float and anchor-weight covered approximately the lower half of the water column.

**Fig. 2**. Five days (12-17 November 2017) of fixed-point observations on a sudden northerly wind event and deep internal wave response in late fall. (a) Along-lake wind components measured at the stations near the northern end of the lake (heavy line) and near the southern end (dashed line), plotted in oceanographic convention. (b) Current-flow components measured near mid-depth at the mooring, with along-lake component in heavy line and cross-lake component in thin line. (c) Time-depth image of potential temperature from the moored temperature sensors. In the very light-grey part the weak stratification yields approximately $N \approx f$. The white bars indicate (from top to bottom): The mean buoyancy period, the minimum buoyancy period and the inertial period. Tick-marks are at semidiurnal periods.

**Fig. 3**. Magnifications from Fig 2c. Note the different greyscale-value ranges between the two panels and between Fig. 2c. (a) Time-depth plot of potential temperature for 3.5 h, 65 m above the lake-floor. The rectangle indicates the magnification of the lower panel. The horizontal axis is at the lake-floor. (b) 0.4 h, 52 m high detail of a. The horizontal scale dX is calculated using the mean mid-depth ($z = -187$ m) flow speed for this time-window.

**Fig. 4**. One inertial period (16.7 h) of moored high-resolution temperature observations with predominantly shear-driven overturning in the weakly stratified layer near the lake-floor. (a) Time-depth plot of logarithm of buoyancy frequency with black isotherms drawn every 0.01°C. The bars on top indicate mean buoyancy period (purple), maximum buoyancy period (light-blue; close to half inertial period) and inertial period (white). The lake-floor

is at the horizontal axis. (b) Time-depth plot of lower 60 m of potential temperature. Isotherms as in a. (c) As b., but for a different colour range highlighting the weakly stratified layer near the lake-floor. Contours as in a. (d) Temperature spectra averaged over the upper 30 sensors (unsmoothed in light-blue, smoothed in red) and over the lower 30 sensors (purple, black). Several frequencies are indicated, including the local inertial frequency f, its higher harmonic 2f, semidiurnal solar $S_2$, and depth-(colour-)specific large-(148.5-m)-scale mean buoyancy frequency N, small-(1.5 m-)scale $N_{max}$, Ozmidov frequency $\sigma_O$ (calculated using mid-depth mean current amplitude). The inertial subrange scaling of $\sigma^{-5/3}$ is indicated by the horizontal solid black line. Several other scalings are indicated by dashed lines and their relative slope value (for a log-log plot): -1/3 (black), +2/3 (blue), +4/15 (green). (e) Coherence for 1.5 m vertical separation distances averaged over all possible independent pairs of T-sensor records in the same vertical ranges and plotted for the same horizontal axis frequency range as for d. The 95% significance level is approximately at Coh = 0.2.

**Fig. 5**. As Fig. 4, but almost one day earlier with predominantly convection-driven overturning in the weakly stratified layer near the lake-floor.

**Fig. 6**. Cartoons of potential generation mechanisms of internal wave induced convection. (a) Internally forced convection via downward acceleration into a near-homogeneous layer preconditioned during the preceding 'upward' wave-phase. Reduced gravity is indicated by g'. (b) Obliquely propagating internal waves (redrawn after [34]).

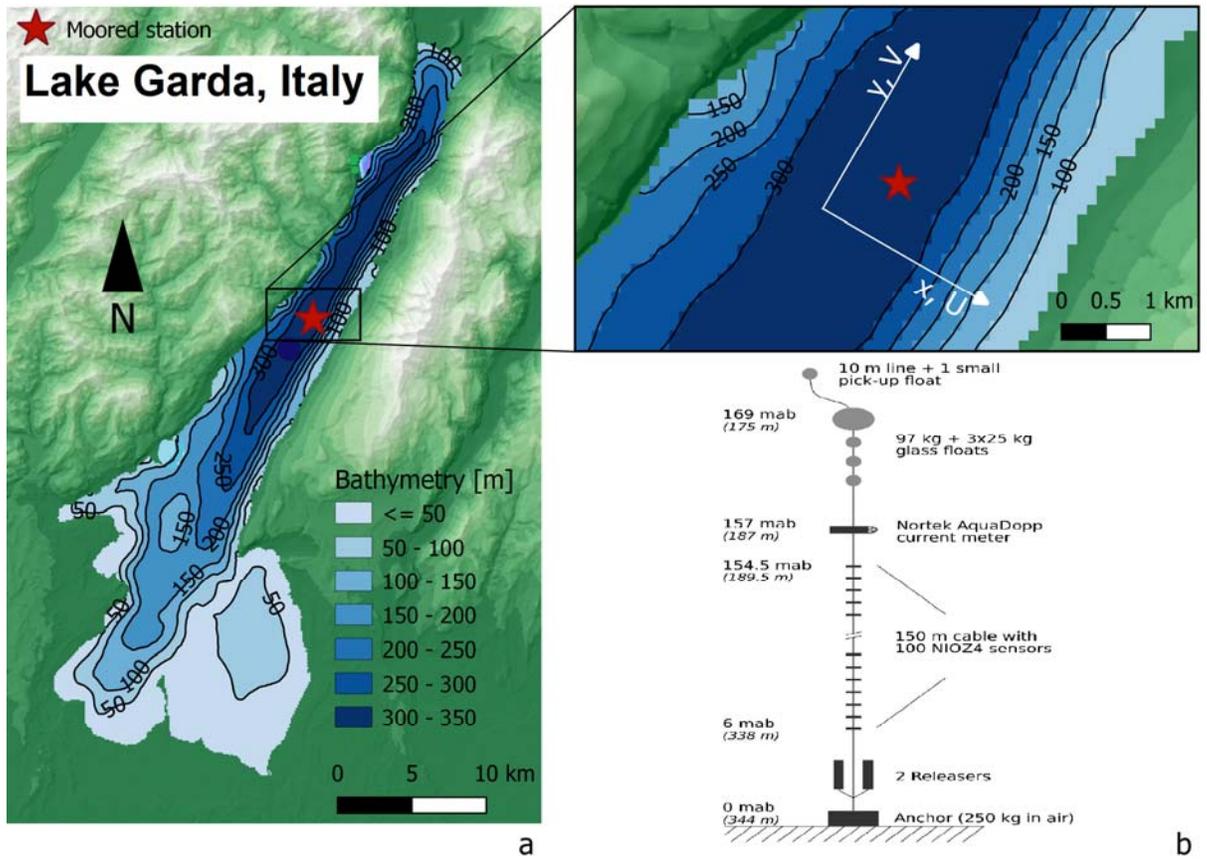

**Fig. 1**. (a) Central part of Lake Garda in Northern Italy with bathymetry, coordinate axes and mooring location (red star) near the lake's deepest point. (b) The instrumented mooring line of about 170 m between top-float and anchor-weight covered approximately the lower half of the water column.

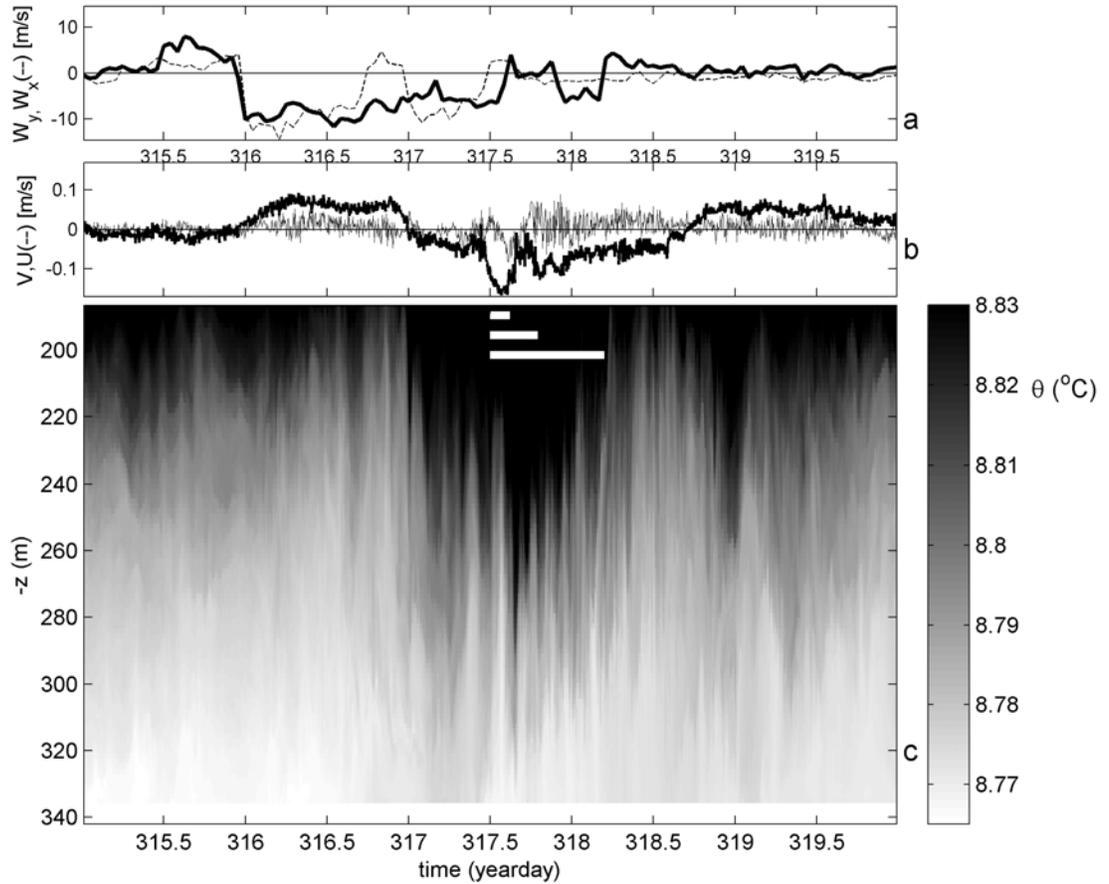

**Fig. 2**. Five days (12-17 November 2017) of fixed-point observations on a sudden northerly wind event and deep internal wave response in late fall. (a) Along-lake wind components measured at the stations near the northern end of the lake (heavy line) and near the southern end (dashed line), plotted in oceanographic convention. (b) Current-flow components measured near mid-depth at the mooring, with along-lake component in heavy line and cross-lake component in thin line. (c) Time-depth image of potential temperature from the moored temperature sensors. In the very light-grey part the weak stratification yields approximately $N \approx f$. The white bars indicate (from top to bottom): The mean buoyancy period, the minimum buoyancy period and the inertial period. Tick-marks are at semidiurnal periods.

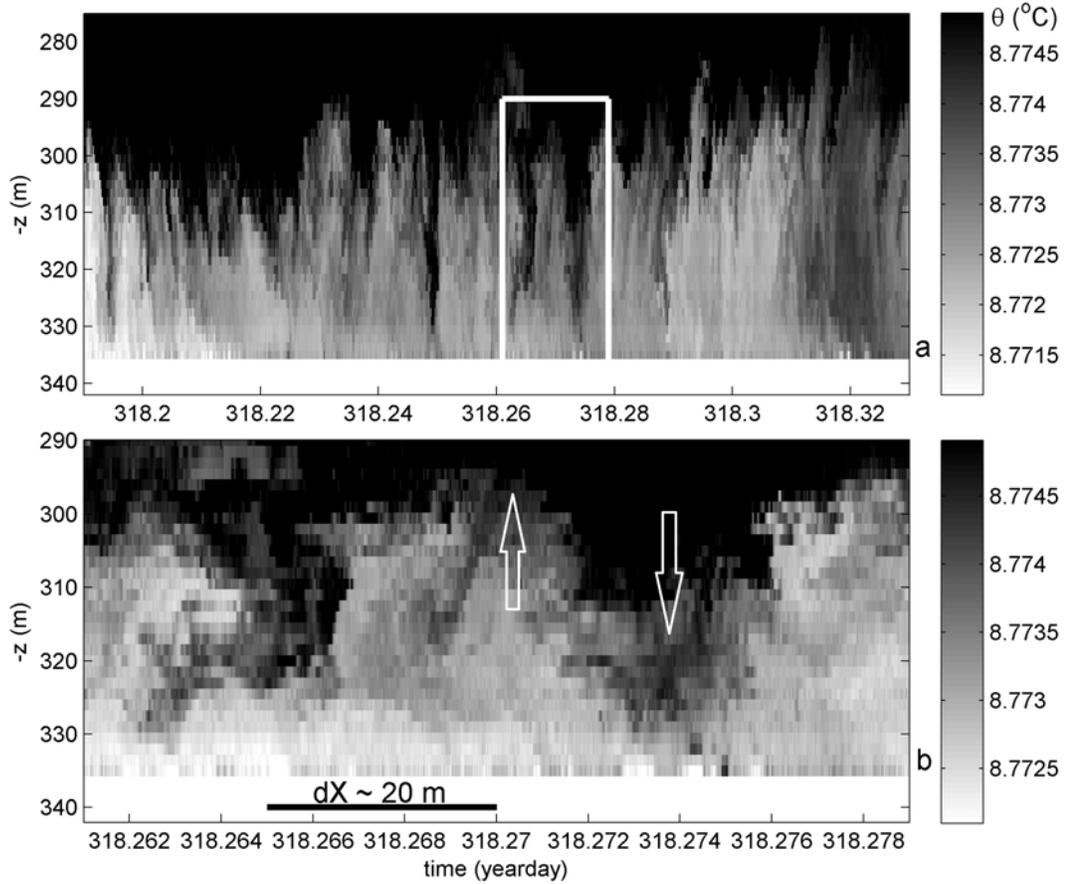

**Fig. 3**. Magnifications from Fig 2c. Note the different greyscale-value ranges between the two panels and between Fig. 2c. (a) Time-depth plot of potential temperature for 3.5 h, 65 m above the lake-floor. The rectangle indicates the magnification of the lower panel. The horizontal axis is at the lake-floor. (b) 0.4 h, 52 m high detail of a. The horizontal scale dX is calculated using the mean mid-depth ($z = -187$ m) flow speed for this time-window.

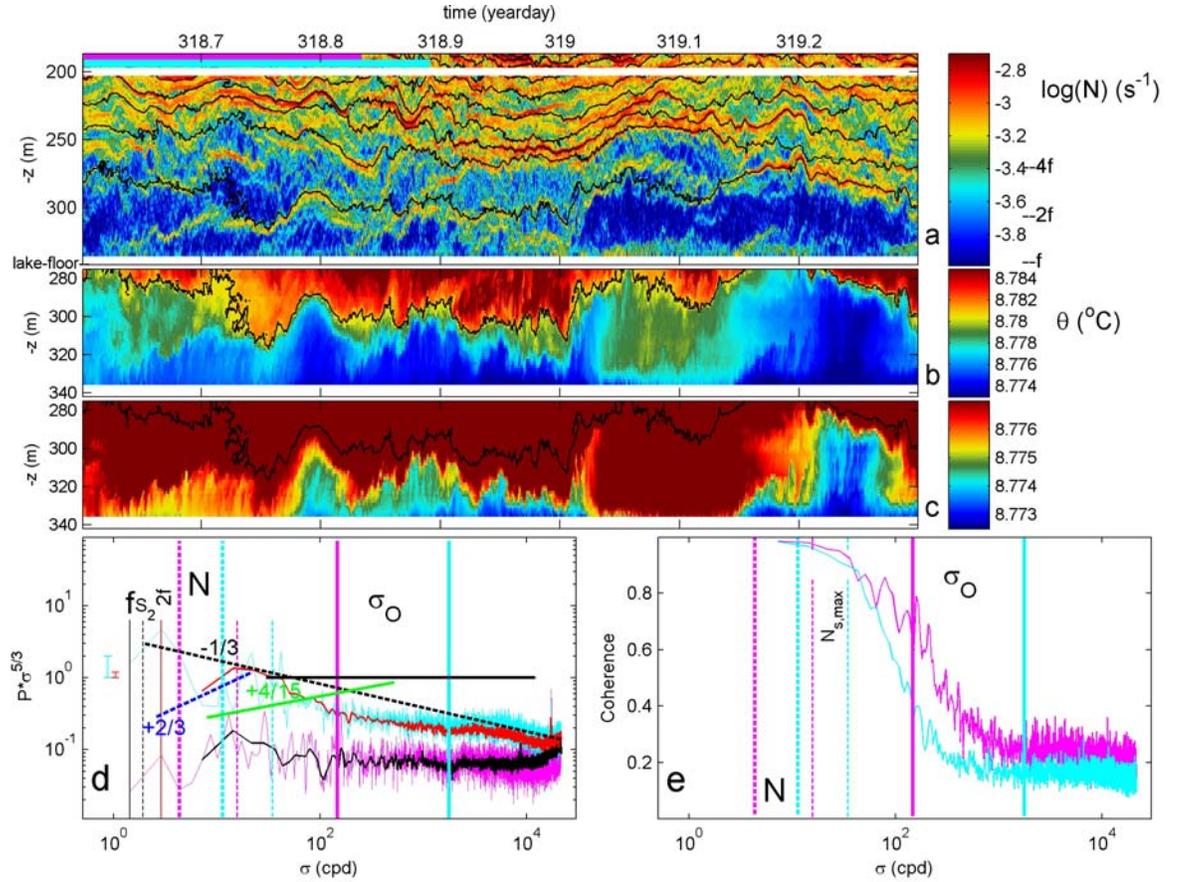

**Fig. 4**. One inertial period (16.7 h) of moored high-resolution temperature observations with predominantly shear-driven overturning in the weakly stratified layer near the lake-floor. (a) Time-depth plot of logarithm of buoyancy frequency with black isotherms drawn every 0.01°C. The bars on top indicate mean buoyancy period (purple), maximum buoyancy period (light-blue; close to half inertial period) and inertial period (white). The lake-floor is at the horizontal axis. (b) Time-depth plot of lower 60 m of potential temperature. Isotherms as in a. (c) As b., but for a different colour range highlighting the weakly stratified layer near the lake-floor. Contours as in a. (d) Temperature spectra averaged over the upper 30 sensors (unsmoothed in light-blue, smoothed in red) and over the lower 30 sensors (purple, black). Several frequencies are indicated, including the local inertial frequency f, its higher harmonic 2f, semidiurnal solar $S_2$, and depth-(colour-)specific large-(148.5-m)-scale mean buoyancy frequency N, small-(1.5 m-)scale $N_{max}$, Ozmidov frequency $\sigma_O$ (calculated using mid-depth mean current amplitude). The inertial subrange scaling of $\sigma^{-5/3}$ is indicated by the horizontal solid black line. Several other scalings are indicated by dashed lines and their relative slope value (for a log-log plot): -1/3 (black), +2/3 (blue), +4/15 (green). (e) Coherence for 1.5 m vertical separation distances averaged over all possible independent pairs of T-sensor records in the same vertical ranges and plotted for the same horizontal axis frequency range as for d. The 95% significance level is approximately at Coh = 0.2.

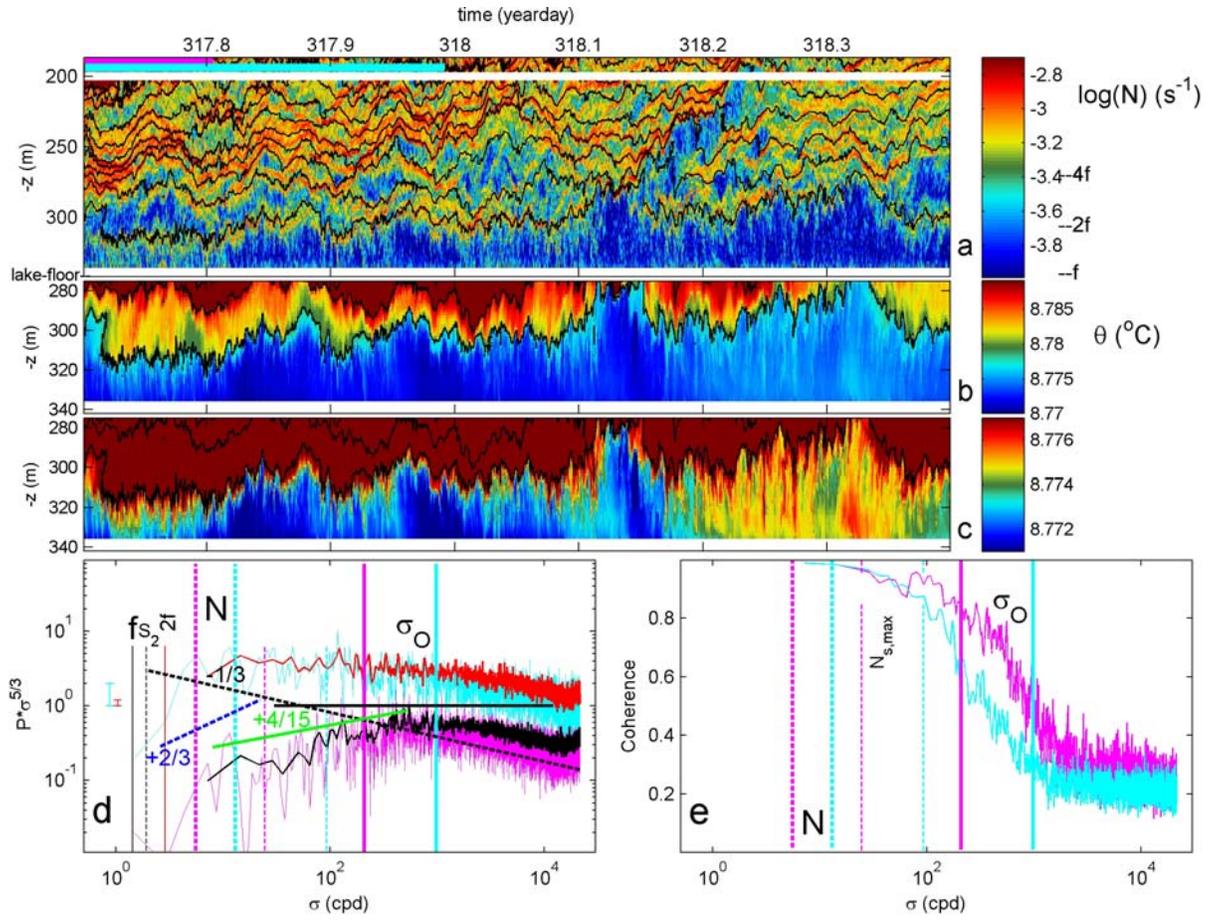

**Fig. 5**. As Fig. 4, but almost one day earlier with predominantly convection-driven overturning in the weakly stratified layer near the lake-floor.

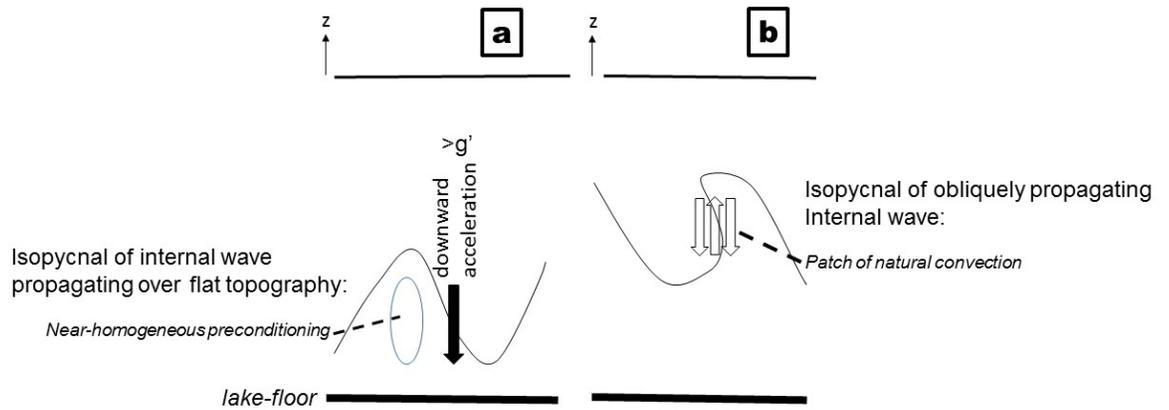

**Fig. 6**. Cartoons of potential generation mechanisms of internal wave induced convection. (a) Internally forced convection via downward acceleration into a near-homogeneous layer preconditioned during the preceding 'upward' wave-phase. Reduced gravity is indicated by g'. (b) Obliquely propagating internal waves (redrawn after [34]).